\documentstyle[aps,twocolumn,epsf]{revtex}
\begin{document}

\title{Reply to ``Comment on ``Some implications of the quantum nature of laser fields for quantum computations'' ''}
\author{S.J. van Enk$^1$ and H.J. Kimble$^2$}

\address{ $^1$Bell Labs, Lucent Technologies, 600-700 Mountain Ave,\\
Murray Hill NJ 07974\\
$^2$Norman Bridge Laboratory of Physics 12-33\\
California Institute of Technology, Pasadena CA 91125}

\maketitle

\begin{abstract}
We point out several superficialities in Itano's comment (quant-ph/0211165).
\end{abstract}

\medskip
Although the recent Comment by Itano\cite{itano} is directed at
two papers by J. Gea-Banacloche \cite{gea}, it also calls into
question our paper \cite{qic}. Here we address the latter
criticisms, and show that the arguments in Itano's comment are
superficial and do not affect the correctness nor conclusions of
our analysis.

Itano claims that a laser field initially in a coherent state does not
become entangled with an atom it is interacting with, in contrast to the
conclusion we reached in \cite{qic}, and that all the decoherence effects
discussed there can in fact be attributed to spontaneous emission.
Three arguments are given for this conclusion: 
\begin{enumerate}
\item The
formalism that we employed in \cite{qic} is ``inappropriate'' for the setting of
free space, since the field is not confined by a cavity.
\item Mollow \cite{mollow} showed that by applying an appropriate
unitary transformation the Hamiltonian can be transformed into one that
describes the interaction of the atom with a classical field and the vacuum.
Clearly, the classical field will not become entangled with the atom, so all
entanglement can only be with the vacuum.
\item In free space, the
atom radiates a dipole field and coherent forward scattering, ``which do
not modify the \textit{incident} field''.
\end{enumerate}
Our responses are
\begin{enumerate}
\item Ref.~\cite{loudon} discusses how one may quantize the electromagnetic field 
in terms of freely propagating modes, not confined by any cavity. 
That is the formalism we used, 
with the propagating laser pulse being one of those modes. 
Now it is true one has to be careful when describing the interaction 
of such a mode with an atom, as pointed out in the paper by 
Silverfarb and Deutsch\cite{deutsch}, with which we agree.
Although Itano refers to the fact   
that Silverfarb and Deutsch have ``independently reached
similar conclusions,'' to those in his comment \cite{itano}, 
this actually refers to another issue:
 Refer
to the second paragraph of Section II in \cite{deutsch}:
``This approach was taken by van Enk and Kimble and also by
Gea-Banacloche ...'' whose ... ``analysis led to an effective single temporal mode theory. 
\textit{Though their conclusions are correct} (our emphasis), one must
take great care to understand the regimes under which this formalism is
applicable, '' which we did indeed do in Ref.~\cite{qic}. 
The subject of our paper was to assess the amount of decoherence due to
the atom-laser field entanglement only,
while leaving out all other decoherence effects, in particular spontaneous
emission into modes that are initially empty. 
We of course agree that if spontaneous emission is included, a single-mode model 
cannot be correct, as explicitly demonstrated in \cite{deutsch}.
The result of \cite{qic} is that
the decoherence effect due to stimulated emission into the laser
mode is much smaller, in general, than decoherence due
to spontaneous emission, but not zero. 
Note that in 
Section III A of \cite{deutsch} it is concluded that 
``decay due to entanglement 
with the laser modes is small compared to decay due to spontaneous emission...'' but this entanglement is not zero, 
exactly as concluded and calculated explicitly in \cite{qic}, but 
in disagreement with Itano's statements, who claims the decoherence 
due to laser-atom entanglement is {\em zero}.
\item The ``vacuum''  in the Mollow
picture is not the standard vacuum. Having initially performed
Mollow's unitary transformation $U$, one has to apply the inverse operation $U^{\dagger }$
 to get back to the correct physical picture. In particular,
if an atom emits a photon into a mode that was occupied prior to the
initial transformation $U$, the ``one-photon
state'' will be transformed by $U^{\dagger }$\ back to a
state that is close to, but not quite equal to, a coherent state. Thus,
the atom becomes entangled with the laser field by stimulated emission into
the laser mode, exactly as we concluded in \cite{qic}.
\item In two previous related papers \cite{focus}, not mentioned by
the Comment, we studied how quantum-statistical properties of an incident
field are modified by its interaction with a single atom. We used a
well-known expression for the total electric field in the Heisenberg
picture, namely $\vec{E}=\vec{E}_{\mathrm{free}}+\vec{E}_{\mathrm{source}}.$
The ``source'' field is a dipole field (in
the far field). If one identifies that field with spontaneous emission
(which Itano seems to do) and the ``free''
field with the incident field, one would indeed conclude that the incident
field is never ever changed (there is only free evolution). While this
may be formally true, it is physically irrelevant, since only the total
field $\vec{E}$ is relevant subsequent to the atom-field interaction. That Itano's
conclusion is odd, to say the least, can be seen from the fact that it would
hold regardless of the state of the incident field, not just for coherent
states, but for Fock states as well. What Itano overlooks is that the
incident laser field will contain dipole waves as well. Subsequent to the
interaction, the dipole waves in the incident field cannot be
distinguished from the dipole waves emitted by the atom. For example, if an atom
scatters a photon from the incident beam into other modes, or if an initially excited
atom deposits a photon into the laser mode, there are
unavoidable imprints of these processes left in the \textit{total }forward
propagating field $\vec{E}$, since after all energy is conserved.
\end{enumerate}
In short, while we agree that one has to take great care applying a single-mode model
to the description of the interaction of an atom with a laser field in free space,
we do not agree with any of the arguments put forward by Itano that purportedly
show that the laser-atom entanglement is zero.

\end{document}